\documentclass[usenatbib]{mn2e}
\usepackage{graphicx,epsfig,natbib,amssymb}
\usepackage{wrapfig}
\def\gtwid{\mathrel{\raise.3ex\hbox{$>$\kern-.75em\lower1ex\hbox{$\sim$}}}}
\def\ltwid{\mathrel{\raise.3ex\hbox{$<$\kern-.75em\lower1ex\hbox{$\sim$}}}}
\def\lessim{\mathrel{\raise.3ex\hbox{$<$\kern-.75em\lower1ex\hbox{$\sim$}}}}
\def\\{\hfil\break}
\def\ie{{\it i.e.\ }}
\def\eg{{\it e.g.\ }}

\def\lesssim{\mathrel{\hbox{\rlap{\hbox{\lower4pt\hbox{$\sim$}}}\hbox{$<$}}}}
\def\gtrsim{\mathrel{\hbox{\rlap{\hbox{\lower4pt\hbox{$\sim$}}}\hbox{$>$}}}}

\newcommand{\mamo}[1]{\mbox{$#1$}}
\newcommand{\unit}[1]{\ifmmode \:\mbox{\rm #1}\else \mbox{#1}\fi}
\newcommand{\mone}{\mamo{^{-1}}}

\newcommand{\kms}{\unit{km~s\mone}}

\newcommand{\mpc}{\unit{Mpc}}

\begin{document}

\title [Unbiased Gaussian Estimator of Peculiar Velocity]
{An Unbiased Estimator of Peculiar Velocity with Gaussian Distributed Errors for Precision Cosmology}
\vskip 0.5cm
\author[Watkins \& Feldman ]{Richard Watkins$^{\dagger,1}$\& Hume A. Feldman$^{\star,2}$ \\
$^\dagger$Department of Physics, Willamette University, Salem, OR 97301, USA.\\
$^\star$Department of Physics \& Astronomy, University of Kansas, Lawrence, KS 66045, USA.\\
emails: $^1$rwatkins@willamette.edu;\, $^2$feldman@ku.edu}

\maketitle

\begin{abstract}

We introduce a new estimator of the peculiar velocity of a galaxy or group of galaxies from redshift and distance estimates.  This estimator results in peculiar velocity estimates which are statistically unbiased and that have errors that are Gaussian distributed, thus meeting the assumptions of analyses that rely on individual peculiar velocities.  We apply this estimator to the SFI++ and the Cosmicflows-2 catalogs of galaxy distances and, using the fact that peculiar velocity estimates of distant galaxies are error dominated, examine their error distributions,    The adoption of the new estimator significantly improves the accuracy and validity of studies of the large-scale peculiar velocity field and eliminates potential systematic biases, thus helping to bring peculiar velocity analysis into the era of precision cosmology.   In addition, our method of examining the distribution of velocity errors should provide a useful check of the statistics of large peculiar velocity catalogs, particularly those that are compiled out of data from multiple sources.

\end{abstract}


\section{Introduction}
\label{sec:intro}

The Doppler effect provides a remarkably accurate method to infer the velocity of a galaxy toward or away from us by measuring the blue or red-shift of its spectral lines respectvely.   However, since  cosmological expansion also causes a redshift, determination of the peculiar (local) motion $v$ also requires the measurement of the galaxy's distance $r$, so that 
\begin{equation}
v= cz - H_o r\,,
\label{eq:vpec}
\end{equation}
where $c$ is the speed of light, $z$ is the redshift, and $H_o$ is Hubble's constant.   This formula assumes a linear Hubble relation.   For more accuracy, particularly at large distances, we can include the effects of cosmic acceleration by replacing $z$ with $z_{mod}$, where
\begin{equation}
z_{mod} = z[1+0.5(1-q_o)z - (1/6)(1-q_o-3q_o^2+1)z^2]\,,
\label{eq:vmod}
\end{equation}
where $q_o$ is the deceleration parameter \citep[see also][]{DavScr14,SprMagColMou14}.   
In addition, we can achieve additional accuracy by accounting for the fact that redshift is not an additive quantity.  Rather than $cz_{mod}= H_or + v$, we instead should write $(1+z_{mod})=(1+H_o r/c)(1+v/c)$, which reduces to the familiar formula at low redshift.   Solving for $v$, we obtain
\begin{equation}
v = \frac{ cz_{mod} - H_or}{1+H_or/c} \approx \frac{ cz_{mod} - H_or}{1+z_{mod}}
\label{eq:releq}
\end{equation}
where in the second expression we replaced $H_or/c$ with $z_{mod}$, a good approximation since the difference between them, which is approximately $v/c$,  is always much less than one.   The second expression is easier to work in practice since it does not introduce new factors of $r$, a quantity that has large uncertainties.

Whereas redshift can be measured very accurately, distance measurements typically have uncertainties of $\simeq 20\%$, so that the uncertainty $\delta v$ in a peculiar velocity is approximately $\delta v \approx 0.20 H_o r$.   Since typical peculiar velocities are thought to be $\simeq 500$ km/sec, we see that for $H_o\approx 70$ \kms \mpc\mone the uncertainties in peculiar velocities become of order their magnitudes for objects at distances $r\gtrsim 35$ Mpc, which includes the region that we would like to use peculiar velocities as a tool to probe large-scale structure.   Thus individual peculiar velocity measurements have very low signal-to-noise, which makes it is necessary to have a large sample in order to extract meaningful information.   

Two major approaches have been used to analyze peculiar velocity catalogs \citep[for reviews see][]{JacBraCiaDav92,StrWil95}.   The first forgoes calculating individual peculiar velocities altogether, and instead uses distance and redshift information to estimate parameters of a model of the peculiar velocity field \citep[for recent usage of this method see][]{NusDav11,CouHofTulGot12,HonSprStaScr14}.  This approach has the disadvantage that it is difficult to quantify how the different parts of the sample volume are contributing to the final parameter estimates.    For example, peculiar velocity samples typically have much more information at small distances than in the outer parts of the sample.   This is due both to the higher density of nearby objects and the higher accuracy with which their peculiar velocities can be determined.   Thus the results of these analyses can end up mostly reflecting the nearby peculiar velocity field.   

The second approach involves combining many individual peculiar velocities into moments of the peculiar velocity field that have much smaller uncertainties, for example, the bulk flow of a sample volume \citep{Kai88,LynFabBur88,CouFabDreWil93,FelWat94,WatFel95,Wil99,CouWilStr00,pairwise00,NusdaCBraBer01,Hud03,pairwise03,HudSmiLuc04,SarFelWat07,WatFel07,FelWat08}.
This approach has the advantage that the contribution of different parts of the sample to a moment can be controlled by using various weighting schemes.   For example, the Minimum Variance method \citep{WatFelHud09,FelWatHud10,AgaFelWat12,WatFel14}  can be used to create moments that probe a volume in a known, standardized way.   It is also straightforward to quantify how these moments probe the power spectrum, making it possible to compare their values to what would be expected, given a particular cosmological model.   

This second approach relies on reducing the errors by averaging over many noisy measurements.  It is important to emphasize that if the distribution of measurement errors isn't symmetric about zero, then the averaging process will result in an incomplete cancellation of the noise leading to a systematic bias, which can suggest an appearance of flows that do not exist, or mask flows that do.      In addition, most analyses make the stronger assumption that the error distribution is Gaussian.  If this assumption is violated, then the validity of the results could be called into question.   

As peculiar velocity surveys become larger and the uncertainties in derived quantities like the bulk flow become smaller, it is important to revisit the validity of our assumptions about the measurement error distribution.   For example, peculiar velocities calculated using the traditional estimator are known to have non-gaussian errors.   This comes about because the quantity estimated in distance determinations is the distance modulus, which is related to the logarithm of the distance.   While distance moduli have Gaussian errors, their exponentiating skews the distribution of the distance errors, hence leading also to skewed errors in the peculiar velocities.   Malmquist bias corrections include correction for this skewness \citep[see, \eg][]{LynFabBur88,FreZehdaC99}; however, as we shall see below, while this correction produces a distribution of errors that is approximately Gaussian, it is not as effective as the new estimator at eliminating skewed tails in the distribution.  
Skewed, non-Gaussian errors invalidate the statistical assumptions of the analysis methods, and further, may lead to biases in the analyses' results and conclusions.   

In this paper we introduce an unbiased estimator of peculiar velocity that has Gaussian distributed errors.   The use of this estimator will greatly increase the accuracy and reliability of any analysis that relies on individual peculiar velocity measurements.   We also examine the statistics of several large-scale peculiar velocity surveys with both our new estimator and the traditional estimator to determine the validity of our assumptions about measurement errors.   

In Section~\ref{sec:method} we describe in detail the peculiar velocity estimator. In Section~\ref{sec:statistics} we discuss the statistics of peculiar velocity surveys. We  conclude in Section~\ref{sec:discussion}.

\section{Peculiar Velocity Estimator}
\label{sec:method}

Our goal is to obtain an estimate $v_e$ of the peculiar velocity of a galaxy or group from the galaxy's redshift $cz$ and an estimate of it's distance $r_e$.   Given Eq.~(\ref{eq:vpec}), the most straightforward estimator is
\begin{equation}
v_e = cz - H_o r_e\,,
\end{equation}
and this is typically the estimator used in peculiar velocity analyses.   However, from a statistical point of view, this estimator has several undesirable qualities. \citep[For a general discussion of the statistics of estimators, see ][]{Lup93}.  First, distance indicators give distance moduli or log distances with Gaussian distributed errors.   Exponentiating skews the error distribution, resulting in distance errors that are not Gaussian distributed.   Second, this estimator is biased in a statistical sense: the average of an ensemble of velocity estimates with different errors is not the true value,  \ie $\langle v_e\rangle \ne v$.  This is the result of the skewness of the distribution of distance errors, which gives rise to $\langle r_e\rangle \ne r$.    These undesirable features can lead to biases in our analyses and in general invalidate our statistical assumptions about the errors in peculiar velocities.   They suggest that we should be investigating other estimators that might be better behaved statistically.   

We propose instead calculating peculiar velocities using the estimator
\begin{equation}
v_e = cz\log(cz/H_o r_e)\,.
\end{equation}
While this estimator may look unfamiliar, it has the statistical properties that we desire in an estimator.   First, since it uses the log distance (or equivalently, the distance modulus), it has Gaussian distributed errors.   It is easy to see that the uncertainty in the peculiar velocity, $\delta v_e$, is given by $\delta v_e = cz\delta l_e$, where $\delta l_e$ is the uncertainty in  the log distance.   Second, we can use $\langle \log (r_e)\rangle = \langle r\rangle$ to show that this estimator is unbiased, as long as the true $v\ll cz$, which is a good assumption for distant galaxies, 
\begin{eqnarray}
\langle v_e\rangle &=& -cz( \langle\log(H_o r_e)\rangle-\log(cz)) \nonumber\\
&=& -cz( \log(H_o r)-\log(cz)) \nonumber\\
&=& -cz(\log(cz-v) - \log(cz)) \nonumber\\
&=& -cz(\log(1-v/cz))\\
&\approx& v\,,
\end{eqnarray}
where we have used Eq.~(\ref{eq:vpec}) to replace $H_or$ with $cz-v$ and we have assumed that the uncertainties in the redshift $cz$ are negligible.   From Eq.~\ref{eq:releq} we see that a more accurate estimator at large redshift is given by
\begin{equation}
v_e = \frac{cz_{mod}}{(1+z_{mod})}\log(cz_{mod}/H_o r_e)\,.
\label{eq:vest}
\end{equation}
with uncertainty $\delta v_e = cz_{mod}\delta l_e/(1+z_{mod})$.
We stress that the assumption that we are making is that the \textit{actual} velocity of the galaxy or group ($v$) is small compared to the redshift, not the estimated velocity ($v_e$).   While estimates of peculiar velocities can be a few$\times10^3$ km/s, it is thought that most actual peculiar velocities are at most a few$\times10^2$ km/s.   Our assumption should hold quite well for galaxies at distances $\gtrsim 20$ Mpc.   As with the traditional estimator, our new estimator can be made more accurate at large distances by replacing $cz$ with $V_{mod}$ given in Eq.~(\ref{eq:vmod}).   

\begin{figure}
  \begin{center}
\includegraphics[scale=0.4]{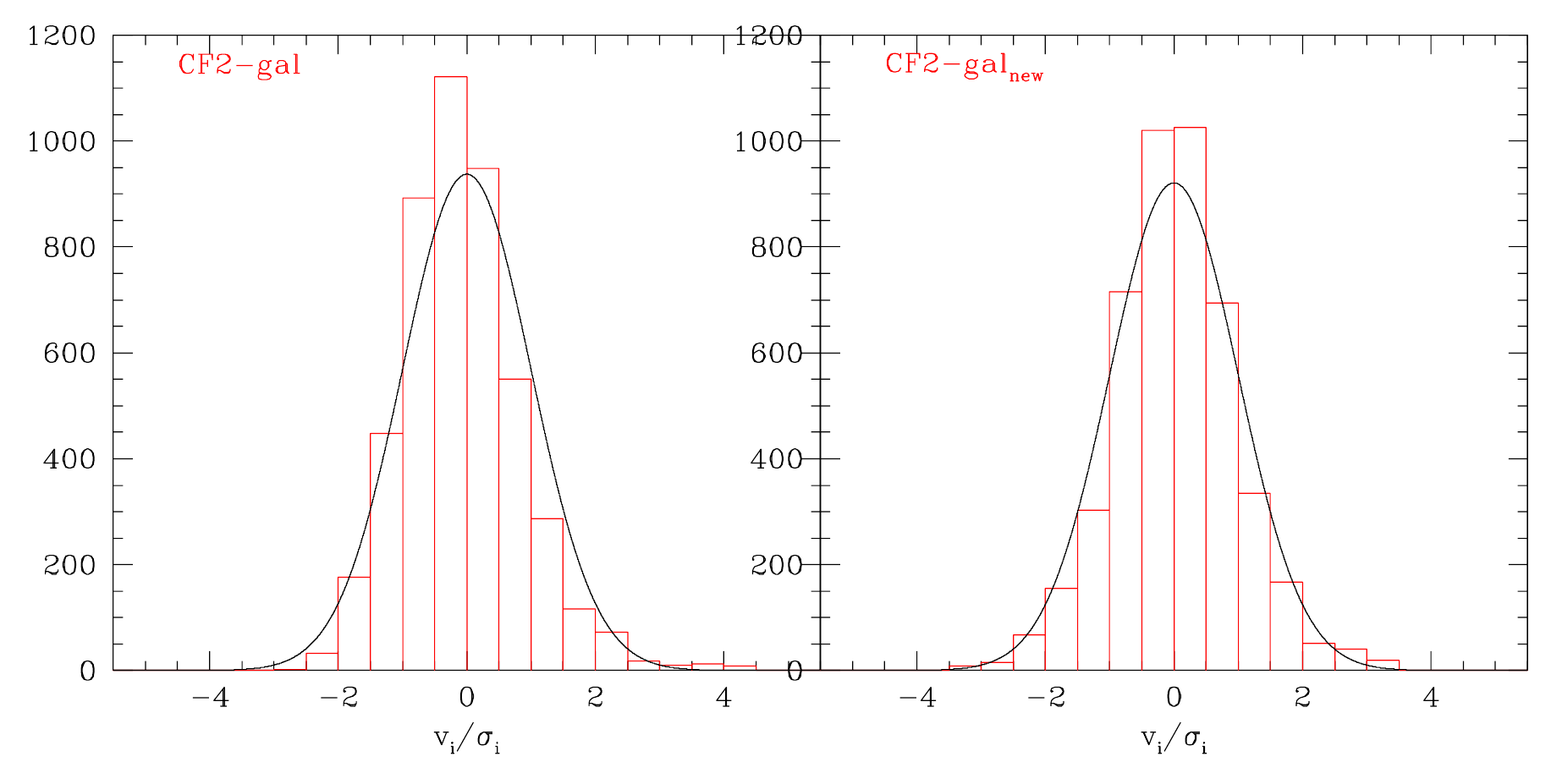}
 \end{center}
 \vspace{-0.5cm}
\caption{\small
The  histograms for the values of the peculiar velocity over their uncertainty, $v_i/\sigma_i$, calculated using both the traditional estimator (left panel) and the new estimator (right panel)  for galaxies with $\sigma_i>1,000$ \kms\, in the CF2 survey. The area of the histograms is normalized to unity. We also show a Gaussian of unit area and unit variance.
}
\label{fig:cf2}
\end{figure}

\section{Statistics of Peculiar Velocity Surveys}
\label{sec:statistics}

The defining characteristic of large-scale peculiar velocity surveys is that they have low signal to noise.   However, if our goal is to determine the distribution of the noise, then we can turn this to our advantage.   In particular, if we consider objects with peculiar velocity errors $\sigma$ such that $\sigma\gg \sigma_v$, where $\sigma_v$ is the spread in actual peculiar velocities, then we can be assured that the objects' measured peculiar velocities are dominated by noise, with negligible contributions from actual motions.    

Here we will examine the error distribution in two large peculiar velocity surveys. The SFI++ \citep{sfi1,sfi2} is a sample of 4052 spiral galaxies with Tully-Fisher distances.   The \textit{Cosmicflows2} (hereafter CF2) \citep{TulCouDol13}  galaxy catalog is a compendium of distances to 8135 galaxies measured with various methods, including Tully-Fisher, Fundamental Plane, SNIa, Surface Brightness Fluctuations, and Tip of the Red Giant Branch.   While the CF2 contains the SFI++ as one of its largest components, in compiling the CF2 a reanalysis of the literature distances was done to ensure consistency between datasets.   For both catalogs we use the more accurate expressions given by Eq.~\ref{eq:releq} for the old estimator and Eq.~(\ref{eq:vest}) for the new estimator, following \citep{TulCouDol13} in assuming the standard cosmological model with $\Omega_m=0.27$ and $\Omega_{\Lambda}=0.73$, so that $q_o = 0.5\left(\Omega_m-2\Omega_{\Lambda}\right)= -0.595$.   

Another difference in the catalogs is that the SFI++  catalog provide distances in \kms and so are scaled relative to the Hubble constant. In contrast, the CF2 sample attempts to determine an absolute scale, and so reports distances in \mpc.  Thus to calculate peculiar velocities from the distances in the CF2 we must assume a value for the Hubble constant.   The nominal value given by the authors of the CF2 in \cite{TulCouDol13} is 74.4 \kms \mpc\mone.   

In Fig.~\ref{fig:cf2} we show histograms for the values of the peculiar velocity divided by their uncertainty, $v_i/\sigma_i$, calculated using both the new estimator and the traditional estimator for galaxies with $\sigma_i>1,000$ \kms\ in the CF2 survey.   If our statistical assumptions are correct, and if actual motions make only a small contribution, then the values $v_i/\sigma_i$ should be unit Gaussian variates, and the histograms should match the Gaussian of unit standard deviation shown in the figure.  We see that the histogram using the new estimator is a good match to the unit Gaussian, but that the traditional estimator results in a skewed distribution.  

\begin{figure}
  \begin{center}
\includegraphics[scale=0.4]{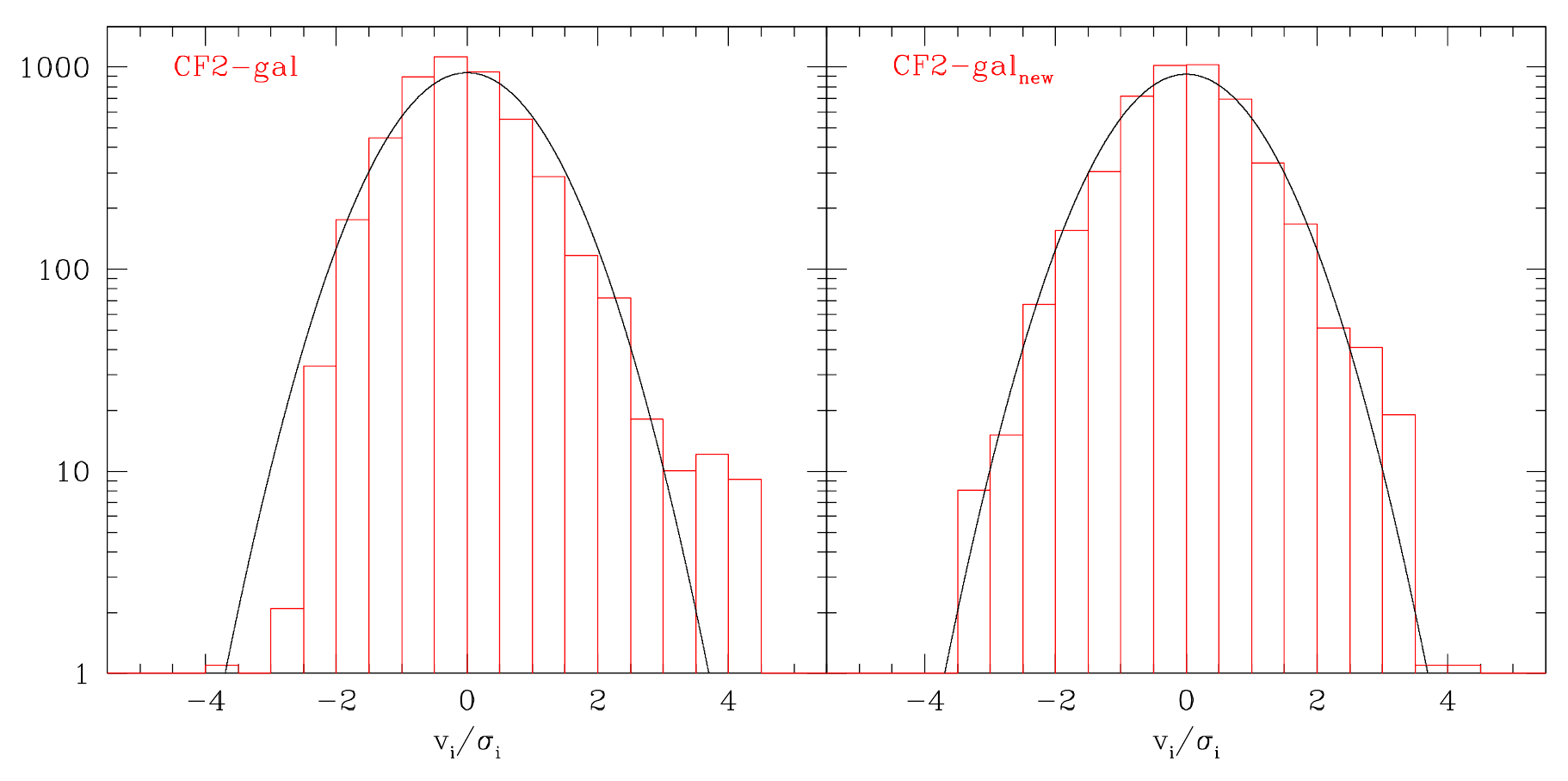}
 \end{center}
 \vspace{-0.5cm}
\caption{\small
Same as Fig.~\ref{fig:cf2} using the logarithm of the histograms to show better the behavior of the tails of the Gaussian distribution.
}
\label{fig:logcf2}
\end{figure}

As seen in the figure, the exponentiation of the Gaussian distributed log distances results in a distribution of errors that is skewed in a complicated way, with the peak shifted toward negative velocities but with a shortened tail on the negative side and an elongated tail on the positive side.   This effect is more clearly seen in Fig.~\ref{fig:logcf2} where we plot the histogram using a logarithmic scale.  This skewness cannot be corrected for by simply shifting the center of the distribution.    Nor can the skewness be corrected by adjusting only negative velocities, as is proposed by  \cite{TulCouDol13} to correct for what they call ``error bias".   

\begin{figure}
  \begin{center}
\includegraphics[scale=0.4]{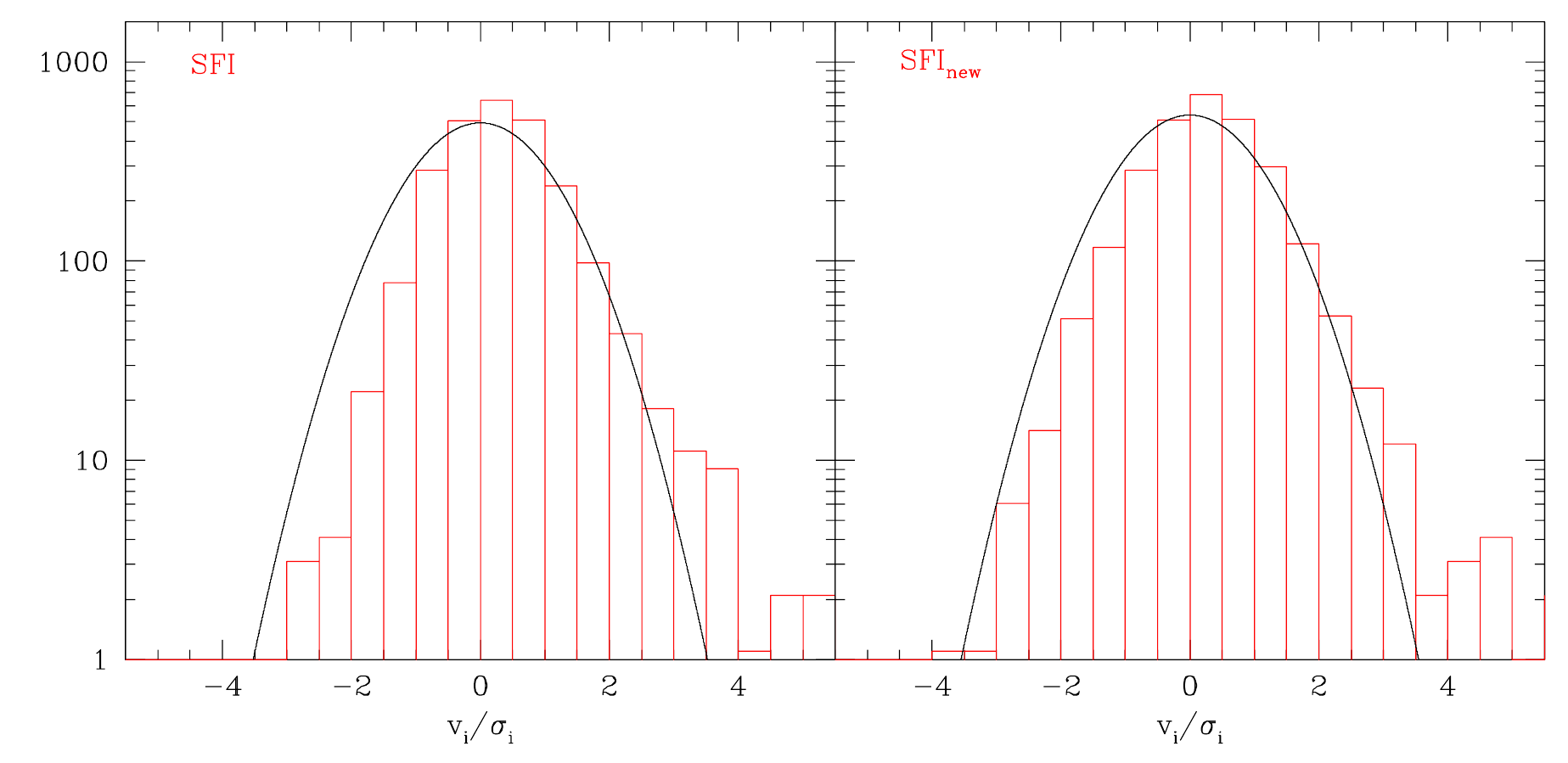}
 \end{center}
 \vspace{-0.5cm}
\caption{\small
The same as Fig.~\ref{fig:logcf2} for the SFI++ survey. We show both the Malmquist corrected traditional estimator (left panel) and Malmquist uncorrected new estimator (right panel) distances.
}
\label{fig:sfi}
\end{figure}

In Fig.~\ref{fig:sfi} we show $v_i/\sigma_i$ histograms for galaxies in the SFI++ survey, again for $\sigma_i > 1,000$ \kms.   As in Fig.~\ref{fig:logcf2}, we plot the histograms on a logarithmic scale to accentuate the tails of the distributions.   The catalogs provide both Malmquist corrected and uncorrected distances.   Malmquist bias correction methods include correction for the skewness of the distribution of distance errors, and \cite{LynFabBur88} showed that velocities calculated with Malmquist corrected distances should be approximately Gaussian distributed.   It does not make sense to use Malmquist corrected distances with the new estimator, since this would in effect double-correct for skewness.  We thus show histograms for the new estimator using uncorrected values and the traditional estimator using corrected values.   First, we see that the histograms for the new estimator and the traditional estimator using Malmquist corrected distances do in fact have the same peak.   However, the histograms are not centered on zero.   

\begin{figure}
  \begin{center}
\includegraphics[scale=0.4]{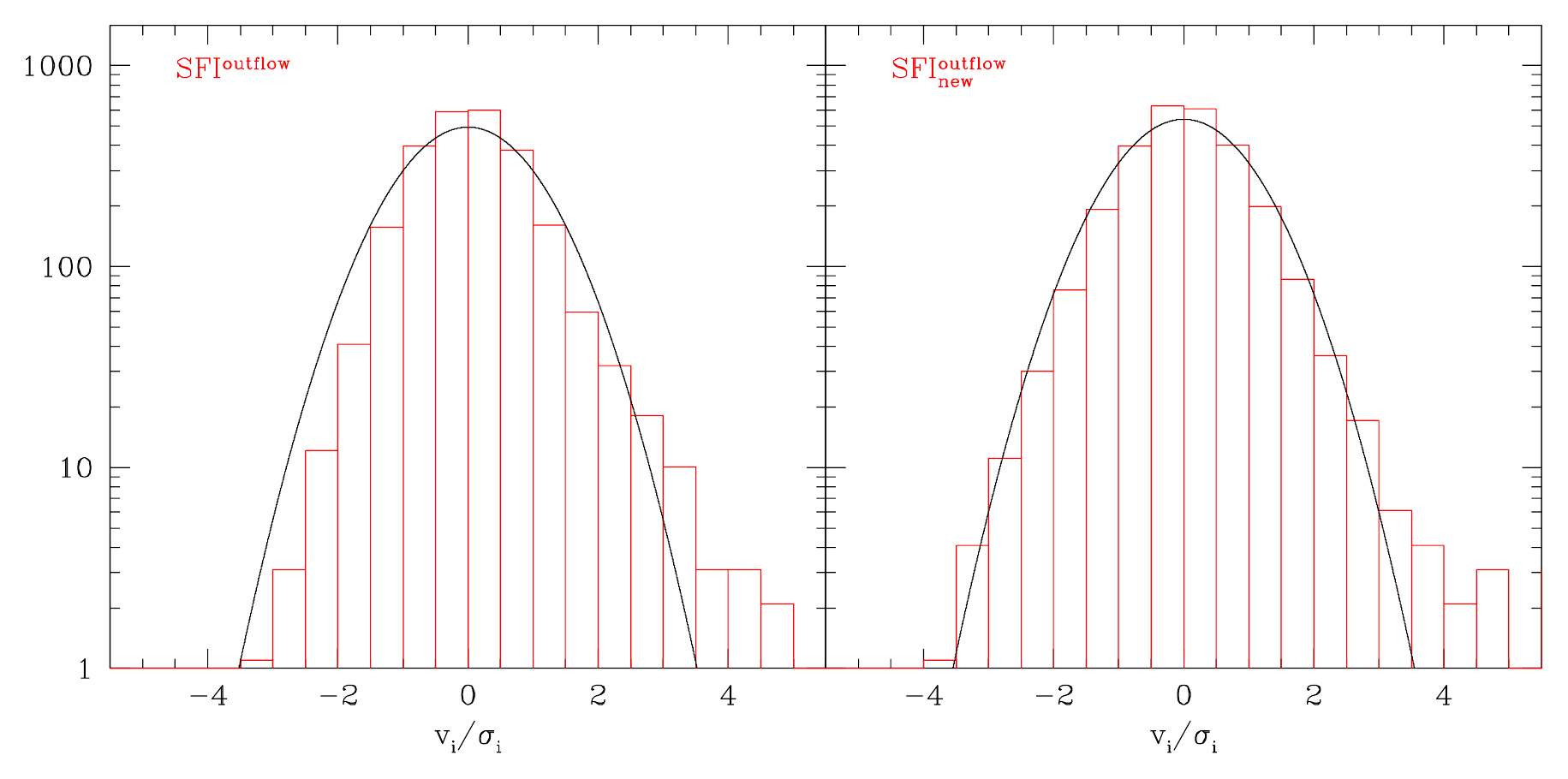}
 \end{center}
 \vspace{-0.5cm}
\caption{\small
The same as Fig.~\ref{fig:sfi} with a 400 \kms coherent outflow in the survey volume added.
}
\label{fig:outsfi}
\end{figure}

One possible explanation for the skewing of the error histograms from zero in Fig.~\ref{fig:logcf2} is a coherent outflow in the volume occupied by the survey.   While random velocities, or even bulk motions, should have little effect on the histograms since their affect would average out over different directions, a coherent outflow would be expected to shift the peak of the histograms toward positive velocities, exactly the effect we see in the figure.   We can test this explanation by considering a simple model 
where the outflow adds a constant peculiar velocity to each galaxy in the survey.   In Fig.~\ref{fig:outsfi} we show the same histograms as in Fig.~\ref{fig:logcf2} except that we have subtracted 400 \kms from the peculiar velocity of each galaxy.   We see that subtracting a relatively modest outflow has resulted in an error distribution that is Gaussian and centered on zero.   The existence of a coherent outflow would support recent arguments that we live in a low density region, dubbed the ``local hole" \citep{WhiSha14,WilSmaMatWat12}.

The disagreement between the CF2 and SFI++ catalogs regarding the existence of a coherent outflow is a consequence of the addition of new data and the reanalysis of literature distances that was done when the CF2 catalog was assembled.   \cite{TulCouDol13} compared distance estimates for galaxies and groups that appeared in more than one component sample and used this comparison to rescale and reanalyze distances to achieve statistical consistency between all the components of the CF2.  This reanalysis was anchored by the zero point provided by Cephied and TRGB distances.  
They found that the resulting CF2 catalog was consistent with a Hubble constant $H_o=74.4 $\kms \mpc\mone\ that did not vary with redshift.   It is worth noting that although this relatively low redshift ($cz\lesssim 0.1$) measurement of $H_o$ is in tension with microwave background results, it agrees well with a recent measurement of $H_o$ using SNIa at much higher redshift \citep{NeiSeiTulCou14}.   

\begin{figure}
  \begin{center}
\includegraphics[scale=0.4]{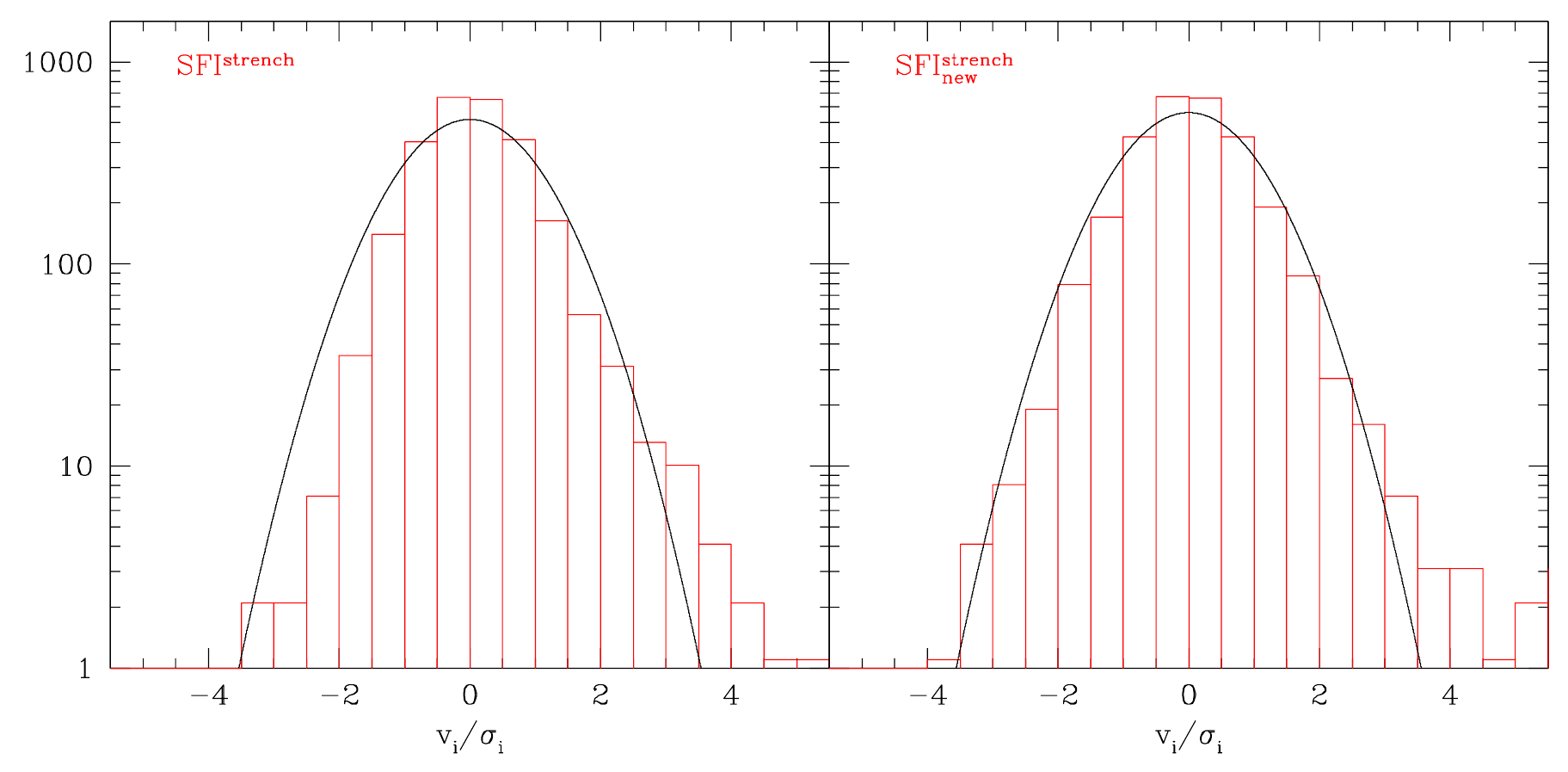}
 \end{center}
 \vspace{-0.5cm}
\caption{\small
The same as Fig.~\ref{fig:sfi} with all distances increased by $5\%$ for the SFI++ survey.
}
\label{fig:sfistretch}
\end{figure}

It is possible that the rescaling of the SFI++ could have inadvertently ``erased" a real coherent outflow.   However, this suggests another explanation of the skewness in the error distribution of the SFI++ survey, a systematic error in the scaling of distances.  In Fig.~\ref{fig:sfistretch} we show the same histograms as in Fig.~\ref{fig:sfi}, but with all distances increased by $5\%$.   Again, we see that there is good agreement between the new estimator histogram and the unit Gaussian centered on zero.   This roughly corresponds to the rescaling of the SFI++ in the CF2.   A direct comparison of the distances given in the SFI++ and the distances for the same galaxies in the CF2, using $H_o=74.4 $\kms \mpc\mone , shows that CF2 distances are about 6.8\% larger on average.   

In both Figs.~\ref{fig:outsfi} and~\ref{fig:sfistretch} we see that although the histogram using the traditional estimator with Malmquist corrected data is indeed approximately Gaussian, the tails of this distribution are still noticeably  skewed.   This demonstrates that our new estimator is more effective at correcting for the skewness of peculiar velocity errors than the method included in Malmquist bias corrections.  



\section{Discussion}
\label{sec:discussion}

Peculiar velocity analysis methods that work with velocity measurements for individual galaxies, groups or clusters assume that the errors in velocity measurements have a Gaussian distribution.   However, the estimator that is traditionally used is known to have a skewed, non-Gaussian error distribution.   
Malmquist bias corrections include a correction that shifts the peak of the error distribution to zero, but these corrections do not remove the skewness in the tails of the distribution.   These tails are particularly important since they represent objects with a large signal to noise ratio.    Since analyses of velocity moments typically weight by uncertainty, objects with a large ratio of estimated peculiar velocity to uncertainty can have a large impact.  

As peculiar velocity catalogs become larger, with a corresponding decrease in the calculated uncertainties in low order moments such as the bulk flow, it becomes increasingly important to address potential systematic errors arising from non-Gaussian velocity error distributions.  
We have introduced a simple, easy-to-use,  peculiar velocity estimator that results in velocities with unbiased, Gaussian errors.   
We have shown that this estimator works well when applied to the CF2 catalog and, with some adjustment, the SFI++ catalog
of galaxy distances.   This new estimator is an important step in bringing peculiar velocity analyses into the era of precision cosmology.    

Our new estimator should not be used to estimate peculiar velocities with Malmquist corrected distances, since currently implemented Malmquist correction procedures include a correction for the skewness of the traditional estimator.  While Malmquist correction does result in approximately Gaussian velocity errors, we have shown that our new estimator does a better job of producing a distribution with symmetric tails.  It is straightforward to modify Malmquist correction methods to be used with the new estimator.   

The new estimator will prove particularly useful for peculiar velocity analyses that are done in redshift space with data that has not been Malmquist corrected, and thus has a velocity error distribution that is biased in addition to being skewed.  For example, the estimator we have introduced alleviates the problem of ``error bias" noted in  \cite{TulCouDol13}.   We have shown that peculiar velocities calculated from CF2 distances using the new estimator have a symmetric, Gaussian error distribution and do not require any further correction.   

We have also presented a method to check large catalogs of peculiar velocities to confirm that they have the expected distribution of errors.   We stress that skewness or non-Gaussianity in velocity error distributions can lead to results which do not accurately reflect the large-scale flows we are trying to study.   This method should provide a useful tool for compiling large peculiar velocity catalogs, particularly when combining data from different sources.  

Finally, we have seen that the distribution of errors in the SFI++ survey is not centered on zero.  This can be explained by an approximately 400 \kms coherent outflow in the survey volume or by a systematic error in the scaling of distances of about 5\%.   Which of these explanations is correct is an interesting question that should be pursued in further research.      

\vspace{0.5cm}
\noindent{\bf Acknowledgements:}  We would like to thank Brent Tully and Mike Hudson for useful comments.


\bibliographystyle{mn2e}
\bibliography{haf}

\end{document}